\documentclass[,final ]
  {aipproc}
\usepackage{extarrows}
\usepackage{subfigure}
\usepackage[breaklinks,hyperindex]{hyperref}
\def\be{\begin{equation}}
\def\ee{\end{equation}}
\def\apj{ApJ}
\def\apjl{ApJL}
\def\aj{AJ}
\def\iaucirc{IAU Circ.}
\def\nat{Nature}
\def\mnras{MNRAS}
\def\physrep{Phys. Rep.}
\newcommand{\etal} {{\it et~al.\ }}

\layoutstyle{6x9}

\begin{document}

\title[Constraints on pulsar evolution]{Do all millisecond pulsars share a common heritage?}



\author{B\"ulent K{\i}z{\i}ltan}{
  address={Department of Astronomy \& Astrophysics, University of California, Santa Cruz, CA, 95064}
}

\author{Stephen E. Thorsett}{
  address={Department of Astronomy \& Astrophysics, University of California, Santa Cruz, CA, 95064}
}

\begin{abstract}
The discovery of millisecond pulsations from neutron stars in low mass X-ray binary (LMXB) systems has substantiated the theoretical prediction that links millisecond radio pulsars (MSRPs) and LMXBs. Since then, the process that produces millisecond radio pulsars from LMXBs, followed by spin-down due to dipole radiation has been conceived as the 'standard evolution' of millisecond pulsars. However, the question whether all the observed millisecond radio pulsars could be produced by LMXBs has not been quantitatively addressed until now. 

The standard evolutionary process produces millisecond pulsars with periods ($P$) and spin-down rates ($\dot{P}$) that are not entirely independent. The possible $P-\dot{P}$ values that millisecond radio pulsars can attain are {\em jointly} constrained. In order to test whether the observed millisecond radio pulsars are the unequivocal descendants of millisecond X-ray pulsars (MSXP), we have produced a probability map that represents the expected distribution of millisecond radio pulsars for the standard model. We show with more than 95\% confidence that the fastest spinning millisecond radio pulsars with high magnetic fields, e.g. PSR B1937+21, cannot be produced by the observed millisecond X-ray pulsars within the framework of the standard model.\footnote{For details see \cite{KT09}: K{\i}z{\i}ltan \& Thorsett (2009)}$^{,}$\footnote{High resolution full color figures available at URL: http://www.kiziltan.org/research.html}
\end{abstract}

\maketitle

\section{Introduction}

Normal pulsars ($P\sim$ few seconds, $B\sim 10^{12}$G) in low mass X-ray binary (LMXB) systems are believed to potentially spin-up to reach milliseconds rotational periods by acquiring angular momentum from their companion \citep{ACR82,RS82}. 

The millisecond modulations in the X-ray signal observed from some of these actively accreting neutron stars and bursting sources are likely to be signatures indicative of their spin periods. There are about $\sim$20 nuclear or accretion powered (see Table~1) millisecond X-ray pulsars (MSXPs) which are thought to be the progenitors of millisecond radio pulsars (MSRPs) \citep{WK98}. These millisecond X-ray pulsars may become observable in radio wavelengths once accretion ceases, or the column density of the plasma from the fossil disk around the neutron star becomes thin enough to allow vacuum gap formation that leads to the production of coherent radio emission. Towards the end of the secular LMXB evolution, as accretion rates fall below a critical value above which detection presumably may be hampered due to absorption or dispersion \citep{TBE94}, the neutron star can re-appear as a millisecond radio pulsar.

Although the connection between LMXBs and MSRPs has been significantly strengthened after the discovery of quasi-periodic kHz oscillations and X-ray pulsations in some transient X-ray sources \citep{WK98, MSS02, GCM02, GMM05}, no radio pulsations from millisecond X-ray pulsars have been detected so far \citep{BBP03}. 

Accretion physics \cite{FKD02} predicts an equilibrium period ($P_{eq}\,\sim\,[\dot{m}/\dot{M}_{Edd}]^{-3/7}$) \citep{BH91} of the neutron star that is accreting in the LMXB system which is constrained by the accretion rate $(\dot{m}$). This equilibrium period combined with the dominant mechanism for energy loss delineates the subsequent kinematics of the spun-up millisecond pulsar. The magnetic dipole model then implies a ``spin-up region'' ($\dot{P}\, \sim\, P_{0}^{4/3} $) \citep[see][]{ACW99} on which the recycled neutron stars will be reborn as millisecond radio pulsars. At the end of the active phase, millisecond X-ray pulsars accreting with $\dot{m}$ and  spinning with $P_{eq}$, presumably transition into a millisecond radio pulsar with an initial spin period of $P_{0} \sim P_{eq}$.

The energy loss due to magnetic dipole radiation after the onset or radio emission is the main process that drives the evolution for millisecond radio pulsars \citep[see][]{MT77, LK04}. Alternative energy loss mechanisms such as multipole radiation or gravitational wave emission, especially during the initial phases of the reborn millisecond pulsars, have been suggested by several authors \citep{K91,B98} but have yet to be observationally corroborated. Advanced Laser Interferometer Gravitational Wave Observatory (LIGO) will be able to probe the frequency space at which millisecond pulsars are expected to radiate gravitational waves, thereby putting stringent constraints on the micro physics of millisecond pulsars.

The advances in radio observations, increased sky coverage with deep exposures of current surveys combined with robust post-bayesian statistical techniques that incorporate minimal assumptions, give us unprecedented predictive power on the joint period-spindown ($P-\dot{P}$) and implied magnetic field ($B$) distributions.

Here, we attempt to go beyond phenomenological arguments and test whether millisecond X-ray pulsars can produce the characteristics of the observed millisecond radio pulsars within the framework of the standard model \citep[and references therein]{BH91}.
	
	\section{\label{sec:dist} Constraining millisecond pulsar evolution}
							
In order to statistically constrain millisecond pulsar evolution, we need to describe the parameters that drive the evolution in quantifiable terms. The evolution of millisecond pulsars can be consistently described in terms of {\bf i)} the equilibrium period distribution ($D$) of millisecond X-ray pulsars at the end of the LMXB evolution {\bf ii)} the mass accretion rates ($\dot{M}$) of the progenitor population during the recycling process  {\bf iii)} Galactic birth rates ($R$), and {\bf iv)} the dominant energy loss mechanism after the onset of radio emission.  

We devise a semi-analytical evolution function $\mathcal{E}$ to parametrize the evolution of millisecond pulsars after the accretion phase, which can be described in closed form as:
\begin{eqnarray}
\displaystyle\sum_{i=0}^{r} \mathcal{E}(D_{i},\dot{M}_{i},R_{i}\, | \, \alpha_{i}^{k},\beta_{i}^{k})  \xrightarrow{n=3} \displaystyle\sum_{i=0}^{r} (P_{i},\dot{P}_{i}) \xrightarrow{} \mathcal{PDF}(P,\dot{P})  \label{stat.eq} 
\end{eqnarray}
where $\mathcal{PDF}$ is the probability distribution function. The shape parameters $\alpha$ and $\beta$ define the distributions (i.e., $D,\dot{M}, R$ for k=1,2,3) for the Beta functions\footnote{Beta functions are commonly preferred in Bayesian statistics as the least restrictive and most flexible prior distributions. It can take the form of an uninformative (e.g. uniform) prior, a monotonic line, concave, convex, unimodal (e.g. normal) or any extreme combinations of these shapes.} \citep{EHP00} inferred from observations at each Monte-Carlo realization ``r''. 

The evolution function $\mathcal{E}$ is built by randomly choosing initiation seeds from a period distribution $D$, which is then convolved via the standard model to consequently sample the $P-\dot{P}$ parameter space. For the observed millisecond X-ray pulsars, the period distribution which seeds will be randomly chosen from is the observed $P_{MSXP}$ distribution (Table~1). We uniquely construct a ``relaxed multidimensional Kolmogorov-Smirnov (K-S) filter'' (Fig.~1) to check population consistencies by calculating the 2D K-S \citep{FF87} probabilities ($P_{2DK-S}$) between observed millisecond radio pulsars and the synthetic population that is formed by these properly evolved progenitor seeds. The filtering is reiterated for each realization to obtain synthetic populations with consistent distributions. The filtered $(D,\dot{M},R)$ values are then convolved with standard dipole energy loss in order to construct the probability distribution function ($\mathcal{PDF}$) which is the integrated form of the proper $(P,\dot{P})$ pairs.

Nominally any $P_{2DK-S} > 0.2$ value would imply consistent populations in a 2D K-S test. By allowing $0.005 < P_{2DK-S} < 0.2$ with lower fractions (see Fig.~1), we oversample outliers to compensate for possible statistical fluctuations and contaminations. A peak sampling rate around the nominal acceptance value of $P_{2DK-S}\sim0.2$ is the most optimal scheme that prevents strong biases due to over or under-sampling. The main goal for oversampling outliers and relaxing the K-S filter is to test whether the standard model can at least marginally produce very fast millisecond pulsars with relatively high magnetic fields like PSR B1937+21.

The predictive significance of the $P-\dot{P}$  distribution for the probability map (Fig.~2) is obtained from a Monte-Carlo run with $r=10^{7}$ valid realizations that produce consistent synthetic samples. Whilst sampling the $P-\dot{P}$ space, no assumptions were made regarding the progenitor period distribution ($D$), the accretion ($\dot{M}$), or the Galactic birth ($R$) rates. The filter (Fig.1) is implicitly driven by the observed millisecond radio pulsars.
	
Figure 2 shows the expected $P-\dot{P}$ distribution for the standard model assuming that millisecond radio pulsars have evolved from a progenitor population similar to the observed millisecond X-ray pulsars. We do not include millisecond radio pulsars in globular clusters because the $P-\dot{P}$ values in these cases may not necessarily be the sole imprint of the binary evolution, but can be significantly changed by possible gravitational interactions due to the crowded field. To explore the extend of the effects of an unevenly sampled progenitor population, we also show the region in the $P-\dot{P}$ space that is sensitive to alternative $P_{MSXP}$ distributions. The probability map is overlaid with the observed millisecond radio pulsars. 

\section{Conclusions}

\begin{itemize}\addtolength{\itemsep}{0.5\baselineskip}
\item  Young millisecond pulsars with higher magnetic fields (e.g. PSR B1937+21) are inconsistent with the standard model. The fraction of the observed young/old millisecond radio pulsars with high B fields is higher than what the standard model predicts by several orders of magnitude. 

\item  The fastest spinning millisecond pulsars, in particular PSR B1937+21, may originate from a different evolutionary channel. 

\item  The standard evolutionary model is able to produce the {\em general} demographics of older millisecond radio pulsars only. 

\item  Accretion rates that millisecond radio pulsars have experienced during their accretion phase deduced from observed $P-\dot{P}$ values, combined with the observed millisecond X-ray pulsar period distribution produces mostly older millisecond radio pulsars, including millisecond radio pulsars with spin-down ages $\tau_{c} > 10^{10}$~yrs. Figure 2 shows clearly that the apparent enigma of millisecond pulsars with spin-down ages older than the age of the Galaxy is mainly a  manifestation of very low accretion rates during the late stages of the LMXB evolution. 

\item  No physically motivated P distribution has been able to produce the whole millisecond radio pulsar population consistently. No millisecond X-ray pulsar period distribution could mimic the observed relative ratios of young/old pulsars with high B fields. 

\item  It is necessary to posit the existence of a separate class of progenitors, most likely with a different distribution of magnetic fields, accretion rates and equilibrium spin periods, presumably among the LMXBs that have not been revealed as millisecond X-ray pulsars. 

\item  A millisecond X-ray pulsar period distribution that has sharp multimodal features coupled with non-standard energy loss mechanisms may be able to reconcile for the joint $P-\dot{P}$ distribution of millisecond pulsars. 

\end{itemize}

\begin{theacknowledgments}
We would like to thank NSHP, NSBP and Vanderbilt University for the invitation and travel support. The research presented here has made use of the 2008 August version of the ATNF Pulsar Catalogue \citep{MHT93}. The authors acknowledge NSF grant AST-0506453.
\end{theacknowledgments}


\bibliographystyle{aipproc}   

\clearpage

\begin{table}
\caption{{\bf Accretion and nuclear powered pulsars}: The millisecond pulsar progenitor seeds used to construct the cumulative synthetic millisecond radio pulsar population for the observed $P_{MSXP}$.}
\begin{tabular}{lcr}
\hline
\hline
{\bf $\nu_{spin} \,[Hz]$} & {\bf Pulsar} & Reference \\
\hline
619\    &\hbox{4U~1608$-$52}   		& 	\cite{HCG03}		 \\
601\    &\hbox{SAX~J1750.8$-$2900}	 &	 \cite{KZH02}		\\
598\    &\hbox{IGR J00291$+$5934} 	& 	\cite{MSS04} \\  
589\    &\hbox{X~1743$-$29} 		          &	 \cite{SJG97}		 \\
581\    &\hbox{4U~1636$-$53}  		&	\cite{ZLS97}	   	\\
567\    &\hbox{X~1658$-$298}  		 &	\cite{WSF01}	         \\    
549\    &\hbox{Aql~X--1}   	 			&	\cite{ZJK98}	         \\
524\    &\hbox{KS~1731$-$260}  		&	\cite{SMB97}	         \\
435\    &\hbox{XTE~J1751$-$305} 		&	\cite{MSS02}	         \\	    
410\    &\hbox{SAX~J1748.9$-$2021}   	&     	\cite{KZH03}	         \\
401\    &\hbox{SAX~J1808.4$-$3658} 	& 	\cite{WK98,CM98}		\\   
377\    & \hbox{HETE~J1900.1$-$2455}	&	  \cite{KMV06}		 \\
363\    &\hbox{4U~1728$-$34}			& 	\cite{SZS96}		  \\
330\    &\hbox{4U~1702$-$429}  		&	\cite{MSS99}		  \\ 		    
314\    &\hbox{XTE~J1814$-$338}		& 	\cite{MS03}		  \\	    
270\    &\hbox{4U~1916$-$05}			&	\cite{GCM01}		  \\
191\    &\hbox{XTE~J1807.4$-$294}		&	\cite{MSS03}  		  \\ 	    
185\    &\hbox{XTE~J0929$-$314}  		&	\cite{GCM02}	       	   \\    
45\     &\hbox{EXO 0748$-$676}   		&	\cite{VS04} 		    \\   
\hline
\end{tabular}
\label{tab:a}
\end{table}
\clearpage

\begin{figure}
  \includegraphics[angle=90,height=.5\textheight]{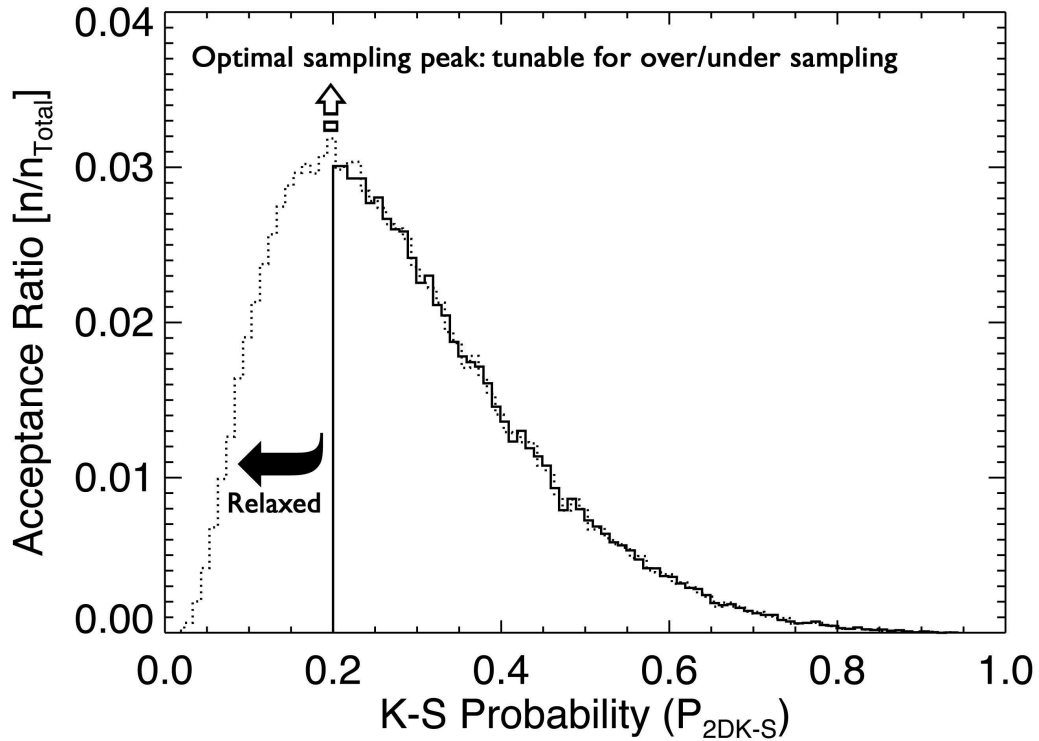}
  \caption{The K-S probability distribution of the synthetic populations used to sample the joint $P-\dot{P}$ parameter space. The relaxed 2D K-S filter allows additional acceptance of populations with $0.05 < P_{2DK-S} < 0.2$ by oversampling the extreme outliers of the $P-\dot{P}$ distribution in order to probe for possible contaminations and extreme fluctuations. The distribution also shows how optimally the $P-\dot{P}$ parameter space is sampled with a peak sampling rate around the nominal acceptance value of $ P_{2DK-S}\sim0.2$. The dotted line is the {\em relaxed} 2D K-S filter for the synthetic populations that is used to construct the predictive distribution in Fig 2. The solid line is the conventional K-S filter that would only accept strictly consistent $P-\dot{P}$ samples.$^{4}$}
\end{figure}
\addtocounter{footnote}{-0}
\stepcounter{footnote}\footnotetext{Labelled version of Fig.1 in \cite{KT09}}
\clearpage

\begin{figure}
	\centering	
	\subfigure[The $P-\dot{P}$ distribution of millisecond radio pulsars evolved from a progenitor population consistent with the {\bf observed} millisecond X-ray pulsars (Table 1). The probability map is a produced by seeds sampled from a {\em discrete} progenitor millisecond X-ray pulsar population.$^{5}$ ]
	{		\includegraphics[angle=90,height=.32\textheight]{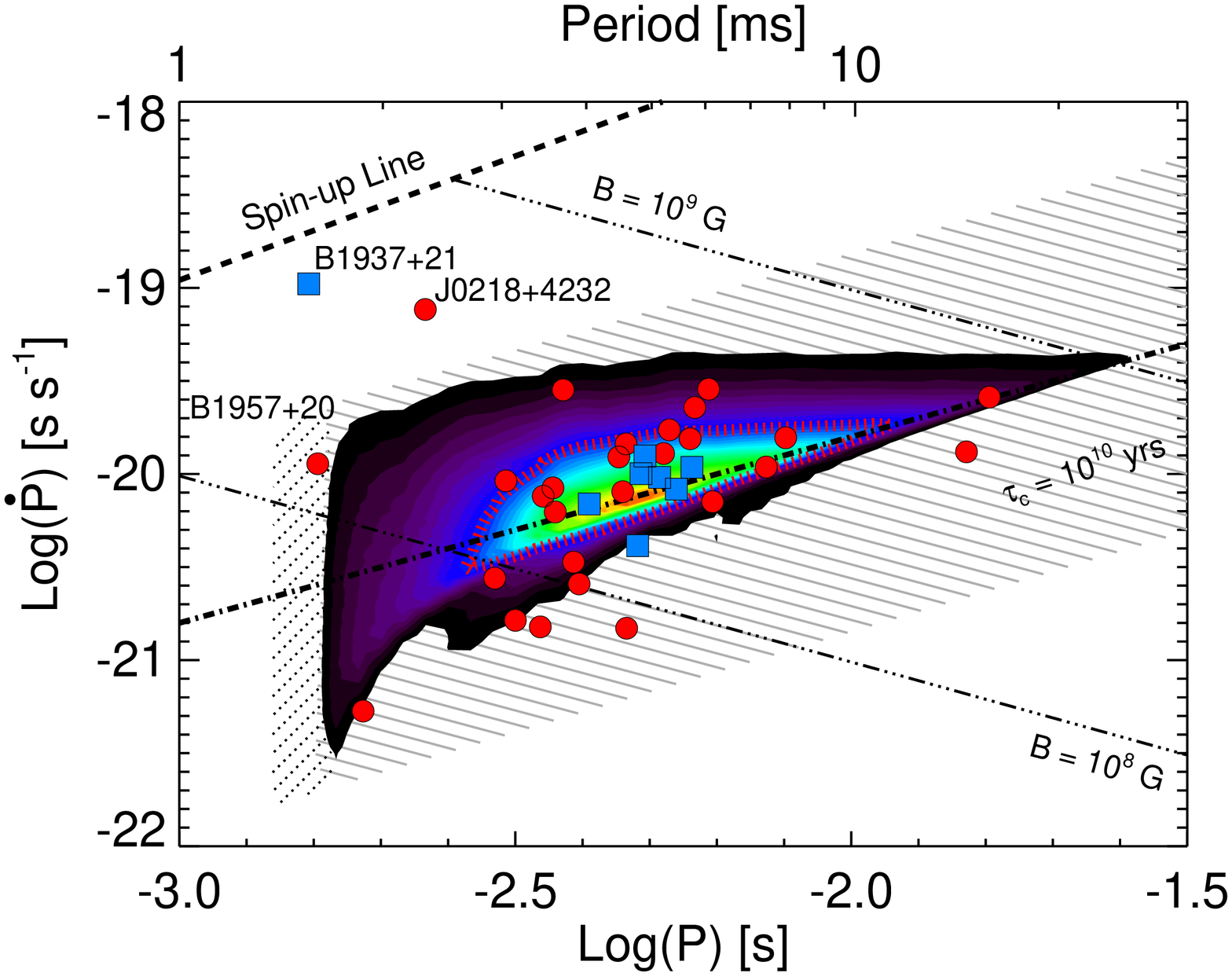} 	}
	 \subfigure[In this case, the probability map is a produced by seeds sampled from a {\em continuous} millisecond X-ray pulsar progenitor population consistent with the {\bf observed} millisecond X-ray pulsars (Table 1).]
	{		   \includegraphics[angle=90,height=.32\textheight]{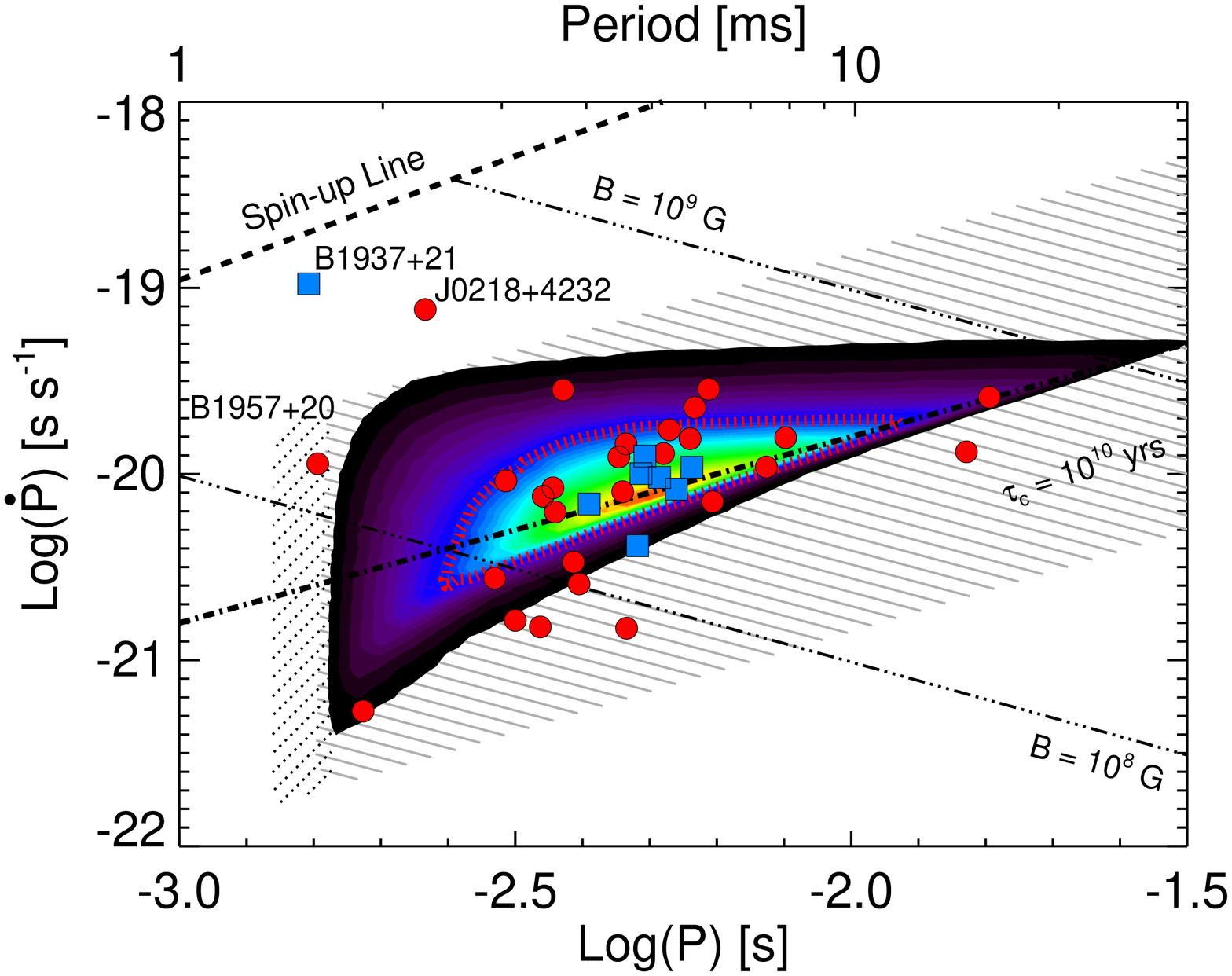} }
	  	 \caption{The $P-\dot{P}$ distribution of millisecond pulsars for the standard model vs. observed millisecond radio pulsars. The color contours are the expected millisecond radio pulsar distribution for the observed millisecond X-ray pulsars, with red representing the highest density. Any millisecond radio pulsar outside of the color contours cannot be produced by the observed millisecond X-ray pulsars with more than 95 \% confidence. The dashed red line is the 68 \%  confidence limit of the expected $P-\dot{P}$ distribution taking the observed millisecond X-ray pulsars as the progenitor population. Millisecond radio pulsars within the shaded region may be produced by millisecond X-ray pulsars with spin distributions different than what is observed. The area shaded with lines assumes a maximum spin frequency $\nu_{max}=619$ Hz, which the fastest spinning observed millisecond X-ray pulsar (i.e. 4U~1608$-$52). The shaded area extends to the dotted region if the maximum spin frequency for millisecond X-ray pulsars is allowed to be $\nu_{max}=760$ Hz which is the theoretical upper limit predicted by \cite{CMM03}. The blue squares and red filled circles are the observed millisecond radio pulsars in single and binary systems respectively. The area outside of the shaded region is not sensitive to the prior, i.e. the observed millisecond radio pulsars outside of the shaded area cannot be produced consistently by the standard model for any millisecond X-ray pulsar spin distribution with more than 95 \% confidence. The spin-down values for the observed millisecond radio pulsars are corrected for secular acceleration \citep{S70, CTK94}. The spin-up line ($\dot{P}\sim \dot{m} P_{0}^{4/3}$) for $\dot{m}=\dot{M}_{Edd}$ and the characteristic age line for $\tau_{c}=10^{10}$ yrs are shown with dashed and dash-dotted lines.$^{6}$}
\end{figure}
\addtocounter{footnote}{0}
\stepcounter{footnote}\footnotetext{Fig.2 in \cite{KT09} optimized for grayscale.}
\stepcounter{footnote}\footnotetext{High resolution full color figures available at URL: http://www.kiziltan.org/research.html}

\end{document}